# Coherent magnon-induced domain wall motion in a magnetic insulator channel

Yabin Fan[1†], Miela J. Gross[2†], Takian Fakhrul[1], Joseph Finley[2,3], Justin T. Hou[2,3], Luqiao Liu[2,3], Caroline A. Ross[1*]

[1]Department of Materials Science and Engineering, Massachusetts Institute of Technology, Cambridge, Massachusetts 02139, USA

[2]Department of Electrical Engineering and Computer Science, Massachusetts Institute of Technology, Cambridge, Massachusetts 02139, USA

[3]Microsystems Technology Laboratories, Massachusetts Institute of Technology, Cambridge, Massachusetts 02139, USA

†These authors contributed equally to this work.

*To whom correspondence should be addressed. E-mail: caross@mit.edu

**Advancing the development of spin-wave devices requires high-quality low-damping magnetic materials where magnon spin currents can propagate efficiently and interact effectively with local magnetic textures. We show that magnetic domain walls (DW) can modulate spin-wave transport in perpendicularly magnetized channels of Bi-doped yttrium-iron-garnet (BiYIG). Conversely, we demonstrate that the magnon spin current can drive DW motion in the BiYIG channel device by means of magnon spin-transfer torque. The DW can be reliably moved over 15 μm distances at zero applied magnetic field by a magnon spin current excited by an RF pulse as short as 1 ns. The required energy for driving DW motion is orders of magnitude smaller than those reported for metallic systems. These results facilitate low-switching-energy magnonic devices and circuits where magnetic domains can be efficiently reconfigured by magnon spin currents flowing within magnetic channels.**

Spin waves, or magnons, can efficiently carry spin information over macroscopic distances in low-damping magnetic materials [1-4]. Magnonic devices are free from certain drawbacks inherent to traditional electronics, such as dissipation of energy due to electric currents [5-7], thus providing a promising platform for realizing ultralow-energy devices and circuits. More importantly, magnons can apply spin-transfer torque [8-10] to a magnetic moment and enable useful functions such as magnetization reversal [11] or domain wall (DW) motion [12,13]. Conversely, magnetic textures such as DWs can reflect, attenuate and shift the phase of a magnon spin current [13-16]. Efficient mutual interactions between magnons and magnetic textures facilitate all-magnon spintronic devices, where magnons can be used to write magnetic bits through magnetization reversal or DW movement and read the bits via their effects on magnon propagation. In prior work, metallic Co/Ni multilayer films with perpendicular magnetic anisotropy (PMA) were utilized to study the interaction between coherent magnons and DWs[13], but the large Gilbert damping ($\alpha = 0.024$) [13] and pinning of the DWs led to low DW velocity (~ 10 m/s) and high microwave power requirements. In contrast, low damping magnetic insulators offer an improved materials platform for wave-based computing by reducing or eliminating dissipation and eddy currents while supporting fast switching dynamics.

In this work, we demonstrate mutual magnon-DW interactions in low-damping bismuth-doped yttrium iron garnet (BiYIG) ferrimagnetic films with PMA: the DW modifies the transmission of spin waves, and the spin waves drive DW depinning and long-range motion even for pulse lengths as low as 1 ns. As schematically depicted in Fig. 1a, spin waves and DWs interact via an exchange of angular momentum. Magnons on the left side of the DW carry spin $+\hbar$ while magnons on the right side carry spin $-\hbar$; when one magnon transits through the DW, a net spin angular momentum of $2\hbar$ is transferred into the DW, which is predicted to drive DW motion opposite to the direction of magnon flow [8-10,17]. In this scenario, similar to current-driven DW motion below Walker breakdown [18-20] the DW velocity is predicted to scale linearly with the magnon spin current [8,12,13]. In contrast, if magnons are reflected or absorbed by the DW, exchange of linear momentum is predicted to drive the DW motion in the direction of the magnon flow, with dynamics that depend on the spin wave frequency, amplitude, and angle of incidence as well as the type of DW and its own spectrum of excitations [21-27]; interfacial symmetry breaking may also affect the direction of spin wave-driven DW propagation [28]. There are few experimental demonstrations of spin wave-driven or -assisted DW motion. Magnon currents originating from a temperature gradient via the spin Seebeck effect have been proposed to explain DW displacement in NiFe[29] and garnet films[12]. Incoherent magnons generated by laser pulses[30] or DW collisions[31] and coherent magnons from an antenna[13] have been used to assist or drive DW motion in metallic films. There are also several examples of DWs modifying the transmission of spin waves in both metallic magnetic films [13,14,16] and in PMA garnet[15], and DWs have been used as conduits, waveguides and magnonic crystals to filter or manipulate the propagation of spin waves [15,22,32-34]. The low-damping BiYIG material employed here allows effective generation of coherent spin waves by an RF antenna at zero external magnetic field, and magnon-driven DW motion using nanosecond RF pulses exhibits orders-of-magnitude lower energy consumption compared with that in metallic systems[13].

The devices in this work were made from a 17 nm thick BiYIG film grown on a $Gd_3Sc_2Ga_3O_{12}$ (GSGG) (111) substrate using pulsed laser deposition (see Methods for details) [20,35,36]. Vibrating-sample magnetometry (VSM) of the as-grown film, Fig. 1b, shows a square hysteresis loop along the out-of-plane direction, with a coercive field $\mu_0 H_c = 0.3$ mT and saturation magnetization of 100 kA/m. High resolution x-ray diffraction (HRXRD) measurements, Fig. 1c, and reciprocal space mapping indicate fully coherent film growth with high structural quality[20]. Field-modulated ferromagnetic resonance (FMR) from 4 to 8 GHz was measured for fields along the out-of-plane direction, Fig. 1d. Zero-field FMR is realized at 3.5 GHz (Fig. 1d), which allows for excitation of coherent magnons while maintaining a specific remanent DW configuration in the device. The Gilbert damping of similar samples 25 nm thick measured in the 10 - 40 GHz range was as low as $\alpha = 1.3 \times 10^{-4}$ (see Supplementary Note 1), consistent with other measurements on BiYIG films[35,37]. This low damping greatly reduces the power dissipation during magnon excitation and propagation compared with metallic films[13].

The BiYIG film was patterned into racetrack structures with width of 5 μm and length of 80 μm, Fig. 2 inset, surmounted by two Ti (10 nm)/Cu (100 nm)/Au (30 nm) microwave antennae for magnon excitation and detection[38,39]. The antennae are U-shaped shorted coplanar waveguides in which the conductors are 1 μm and 2 μm wide while the gap is 1 μm [38]. The two antennae are placed 4 μm (as in Fig. 2 inset), 10 μm, or 15 μm apart. The antenna design allows for the generation of coherent magnons over a range of GHz frequencies, differing from narrow-band magnons generated by DC spin injection[4,40] or non-coherent magnon excitation by a thermal gradient[12] or spin-Hall effect[41].

The propagation of magnons between the two antennae with and without the presence of a DW in the gap is studied by integrating the microwave circuits with a magneto-optical Kerr effect (MOKE) microscope[13]. In the setup (Fig. 2 inset), the magnon injector antenna was connected to a signal generator while the detector antenna was connected to a RF diode detector. The transmission of magnons was measured using a field modulation technique by placing the modulation coil under the device chip (see Supplementary Note 2). The magnetization state of the BiYIG channel was set by applying millisecond external magnetic field pulses and was visualized from the MOKE contrast.

In Fig. 2, the transmitted microwave signal is plotted versus the microwave carrier frequency for remanent states without and with a DW between the two antennae. In the case without a DW, a maximum in transmission was centered at 4.1 GHz, with a peak width of 0.3 GHz. The magnetic nature of the observed resonance can be confirmed by applying a DC external out-of-plane field $H_{ext}$, which shifts the resonance frequency $f$ (see Supplementary Note 3). In the presence of a DW, a resonant signal (appearing at negative voltage due to the reversal of the magnetization) was measured again at 4.1 GHz, with a width of 0.1 GHz. Comparing these two cases, we see that the resonance frequency is maintained but the signal amplitude is attenuated by a factor of 2.4 (or a power reduction by a factor of 5.8) after traveling through the DW. The attenuation in transmitted magnon signal is consistent with previous results from a Co/Ni multilayer with PMA[13].

To demonstrate magnon-induced DW motion we first prepare a DW near one antenna by applying external magnetic field pulses. In Fig. 3a1, a DW (boundary between gray-dark regions indicated with a red arrow) is formed 15 μm from the lower antenna. After the application of one RF microwave pulse of (14 dBm, 1 ms) at 4.1 GHz onto the lower antenna, the DW moves by 3 μm toward the antenna; a second RF pulse of (15 dBm, 1 ms) moves the DW all the way (by 12 μm) to the lower antenna edge. 4.1 GHz was chosen based on its high transmission amplitude shown in Fig. 2. In comparison, when the device was driven by off-resonance 2 GHz microwave pulses at the same power, no DW motion was detected (see Supplementary Note 4). Under the 4.1 GHz signal the DW moves toward the magnon source which is consistent with the magnonic spin-transfer torque mechanism resulting from magnon transmission through the DW [8-10,42], and different from both linear momentum transfer [21] and from field-induced DW movement in the presence of transient spin waves [31], where in both cases the DWs move in the direction of magnon propagation.

We then study the motion of the DW in the region of the spin wave channel between two antennae, reflecting a more practical configuration for programing a spin wave conduit using a magnon current. We note that DWs were commonly found near the edges of antennas, possibly because of local changes in anisotropy due to, *e.g.*, strain from the overlaid metal feature. Fig. 3b1 shows a device where the two antennae have a separation of 10 μm. A DW was formed near the upper antenna edge with external magnetic field pulses, as indicated by the gray-white boundary and red arrow. By applying an RF pulse (2 dBm, 1 ms) at 4.1 GHz to the lower antenna, the DW moves by 2 μm toward the lower antenna, Fig. 3b2. Subsequent pulses at (3 dBm, 1 ms) then (4 dBm, 1 ms) move the DW 1 μm then 7 μm, Fig. 3b3 and 3b4, towards the lower antenna. A local resistance measurement of the antenna in Figure 4 while applying a 10-second 4.35 GHz pulse at 24 dBm showed a temperature increase of 0.07 K (see Supplementary Note 5), which excludes possible Joule heating-induced DW motion.

We explore the effects of incoming RF pulse power, width, and frequency on DW motion in a third device with 15 μm separating the two antennae. We measured the RF power threshold required to initiate DW motion as a function of the carrier frequency of the pulse and the pulse width. Fig. 4a plots the power threshold as a function of carrier frequency for pulse widths of 1 ns, 2 ns, 3 ns, and 75 ns. These data exhibit a minimum power threshold at 4.35 GHz for the device under test. A device with 10 μm separating two antennae, Fig. 3b, gives similar conclusions but with a minimum power threshold at 4.1 GHz (see Supplementary Note 6). The antenna design used here can produce an RF field over a range of frequencies, so the minimum power threshold frequency is not attributed to the antenna design. Instead, the frequency-dependence of threshold power implies that the transfer of energy from the RF field into spin waves, and/or the spin wave-induced excitation, depinning and motion of the DW are most efficient at a particular frequency. Variability in the optimum frequency between devices likely reflects differences in their magnetic properties or pinning sites due to variations in deposition or patterning.

Figure 4a inset shows that an RF pulse as short as 1 ns (19 dBm) at 4.35 GHz caused the DW to move the entire 15 μm distance between the antennae. From the smallest measured power threshold for a 1 ns pulse width, we may estimate the energy required to translate a DW in the channel by estimating losses in the cables and connectors, the inefficiency of energy conversion from the RF signal into spin waves, as well as the spin wave energy dissipated by damping (see Supplementary Note 8), yielding an order of magnitude estimate that the upper bound of energy used to propagate the DW 15 μm is ~ 4 pJ. This energy is orders of magnitude lower than that reported in the Co/Ni metallic multilayer[13] where the energy to drive DW motion is on the order of ~μJ, likely due to the low-damping, -pinning and -dissipation characteristics of the insulating BiYIG.

We now consider the effect of pulse width on the power threshold at the optimum carrier frequency of 4.35 GHz for the device with 15 μm antenna separation, Fig. 4b. The power threshold generally increases as the pulse width is reduced. In the short-pulse regime (< 30 ns), the slope of threshold power vs. pulse width fits well to a model of constant energy (orange dashed line). In this regime, the energy transferred from the spin wave to the DW both depins and propagates the DW. Recalling that spin wave at a frequency $\omega$ has a power $\hbar\omega n v_{\text{SW}} A$ with $n$ and $v_{\text{SW}}$ being the density and velocity of magnons and $A$ the spin wave channel cross-section area, a constant energy therefore indicates that the total number of transferred magnons within the pulse width $\Delta t$, $nv_{SW}A\Delta t$ is a constant, consistent with a fixed total angular momentum required for domain switching. For longer pulses (≥ 1 μs), the threshold power decreases slowly with increasing pulse length. We attribute the behavior in the long-pulse regime to stochastic depinning of the DW at room temperature, in which spin wave excitation of the DW acts to lower the energy barrier for depinning through magnonic spin torque. If we fit this regime using an Arrhenius law [43] (see Supplementary Note 9) we find the energy barrier that pins the DW to be (0.62 ± 0.01) eV. That may be compared to values of 1.379 eV and 1.7 eV obtained in permalloy nanowires and 1.44 eV for FePt [43,44]. The low magnetization, coercivity and damping of BiYIG may account for its lower pinning energy barrier compared to these metals.

**Discussion**

In this work, we have demonstrated mutual interactions between magnons and DWs in channel devices made of a low-damping, insulating, PMA thin film material, BiYIG. We show that the DW attenuates magnon transport. More importantly, we have shown that a magnon spin current generated by an RF pulse applied to an antenna can reliably drive DW motion in a channel device, with a displacement of as much as 15 μm resulting from a 1 ns pulse and an energy requirement in the pJ range. The low Gilbert damping of BiYIG allows for long magnon decay lengths and the insulating properties of BiYIG further reduce Joule heating or conduction electron-related dissipation, making this material ideal for low energy wave-based computing. Unlike spin orbit torque-driven DW motion in Pt/BiYIG[20], the magnon-driven DW motion in the BiYIG channel device does not require the assistance of any external magnetic field to form Néel type DWs[20,45]. To further reduce the power consumption, the antennas may be replaced by a more efficient magnon generation process, for example using DC spin-Hall effect induced auto-oscillations[46,47]. The high DW velocity and ultralow power consumption enabled by low-damping PMA BiYIG facilitates applications in magnon-based spintronic devices and circuits where DWs and other magnetic textures may be effectively written, reconfigured, and detected by a magnon current.

**Methods**

**Material growth:** 17 nm thick BiYIG films were grown by pulsed laser deposition at 560 ˚C substrate temperature and 150 mTorr oxygen pressure from a $Bi_{0.8}Y_{2.2}Fe_5O_{12}$ target onto a $Gd_3Sc_2Ga_3O_{12}$ (GSGG) (111) substrate[20,36]. The 248 nm KrF Compex excimer laser was operated at 400 mJ/pulse (fluence of ~1.9 $J/cm^2$), 10 Hz laser repetition rate, and 6 cm target-substrate distance. The chamber was pumped to $6 \times 10^{-6}$ Torr prior to introducing oxygen. The films were cooled to room temperature at 10 ˚C/min in 230 Torr oxygen.

**Integrated microwave and MOKE setup:** The BiYIG device was attached to a coplanar waveguide printed-circuit board (PCB) with a central hollow area. The PCB had two RF connectors on the two ends. The attennae on the device were connected to the signal line and ground on each side of the PCB waveguide using wire-bonding. The PCB was mounted onto a circular modulation coil and the assembly was installed onto a MOKE microscope stage. Another perpendicular electromagnet was installed on the MOKE microscope from underneath and its pole was extended through the modulation coil to approch the underside of the PCB. This perpendicular electromagnet applied a DC magnetic field to the device to initialize the domain configuration. The two RF connectors on the PCB were connected to a RF signal generator or RF diode by coaxial cables for the microwave experiment. More details of the integrated microwave and MOKE setup can be found in Supplementary Note 2.

**Microwave pulse experiment:** To generate microwave pulses with width longer than 50 ns, we utilized the intrinsic pulse generation function within the Anritsu 68369A/NV synthesized signal generator. For microwave pulses shorter than 25 ns, we combined the Anritsu signal generator with an arbitrary-pulse generator, Hewlett Packard 8131A, through a multiplexer to multiply the nanosecond pulse function from the arbitrary-pulse generator with the microwave signal from the Anritsu signal generator, and characterized the microwave pulse profile using an oscilloscope. The microwave pulse magnitude generated in both cases was calibrated by an Anritsu spectrum analyzer, and loss from both setups was calibrated by a vector network analyzer.

**Acknowledgements**: The authors acknowledge support from SMART, one of seven centers of nCORE, a Semiconductor Research Corporation program, sponsored by the National Institute of

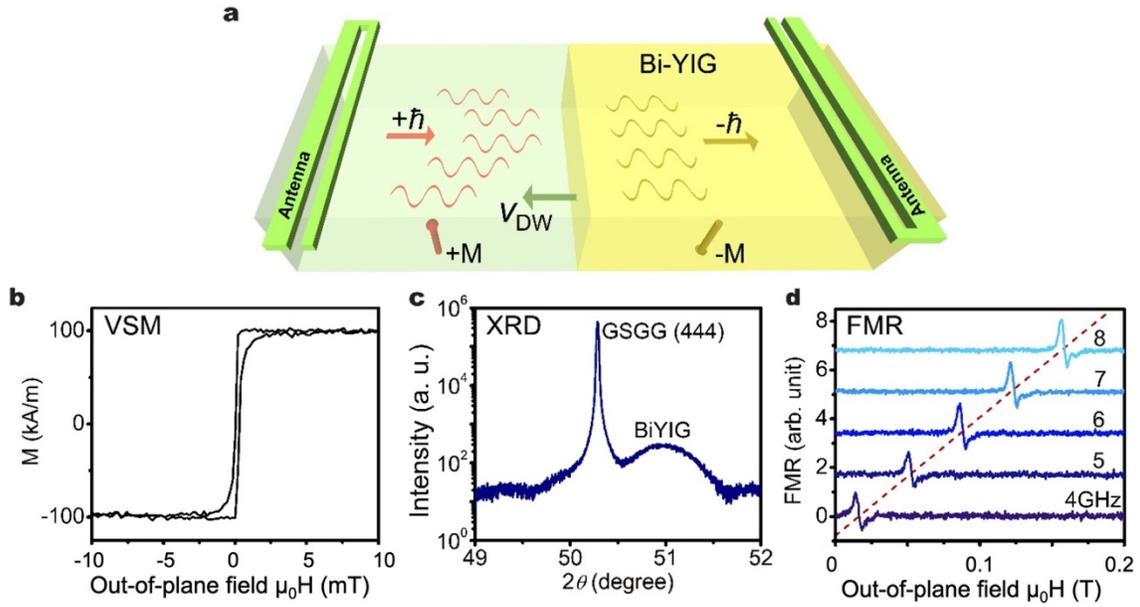

**Fig. 1 | Schematic of magnon-DW mutual interactions and characterization of the Bi-YIG thin-film sample. a**, Illustration of magnons transmitting through a DW in BiYIG. The magnon on the left side of the DW carries spin $+\hbar$ while the magnon on the right side of the DW carries spin $-\hbar$; when a magnon travels through the DW from left to right, a net spin angular momentum of $2\hbar$ is transferred to the DW, leading to its motion with speed $v_{\text{DW}}$. **b,** VSM measurement of the 17 nm BiYIG thin-film sample along the out-of-plane direction. **c**, XRD measurement on the BiYIG film. **d**, Field-differential FMR measurement on the Bi-YIG film sample under different frequencies, with the external field applied along the out-of-plane direction.

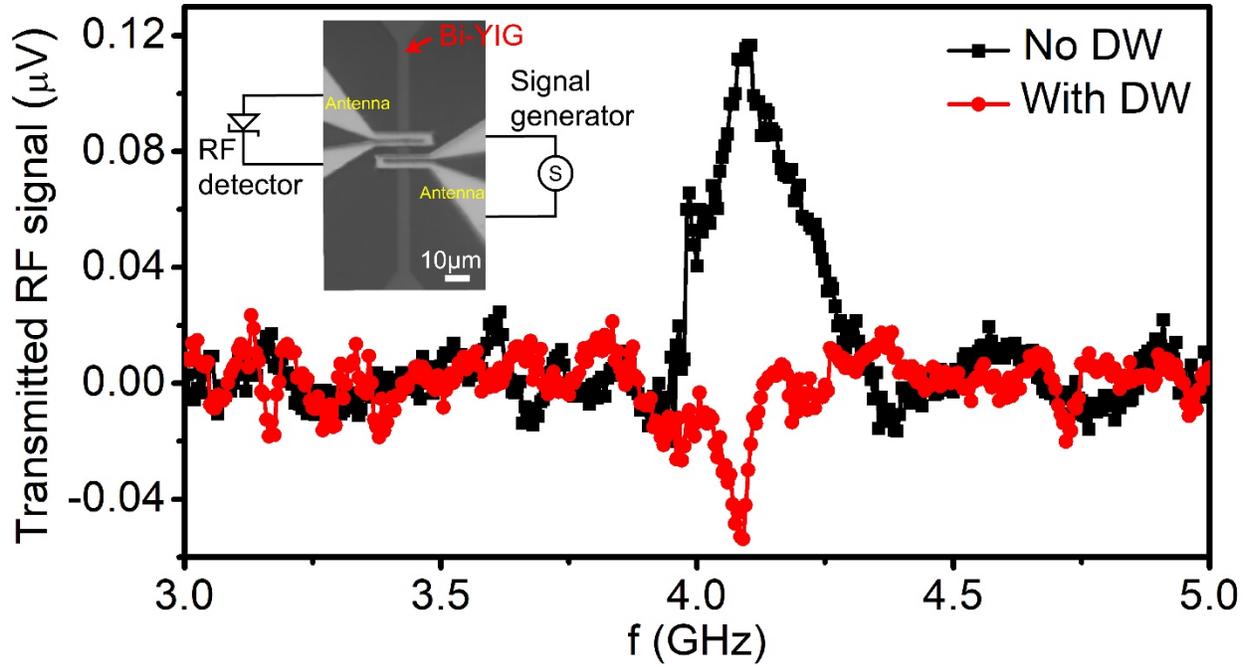

**Fig. 2 | Magnon transmission spectrum of the Bi-YIG channel device with and without the presence of a DW.** Field-modulated magnon transmission spectrum measured for a device with uniform magnetization and with a DW in the gap between the two antennae. The RF power used is 14 dBm. Inset: Optical image of the 5 μm wide BiYIG channel device and illustration of instrument connection in the magnon transmission experiment. Two antennae, one injector and one detector, are fabricated on top of the magnetic channel, separated by 4μm.

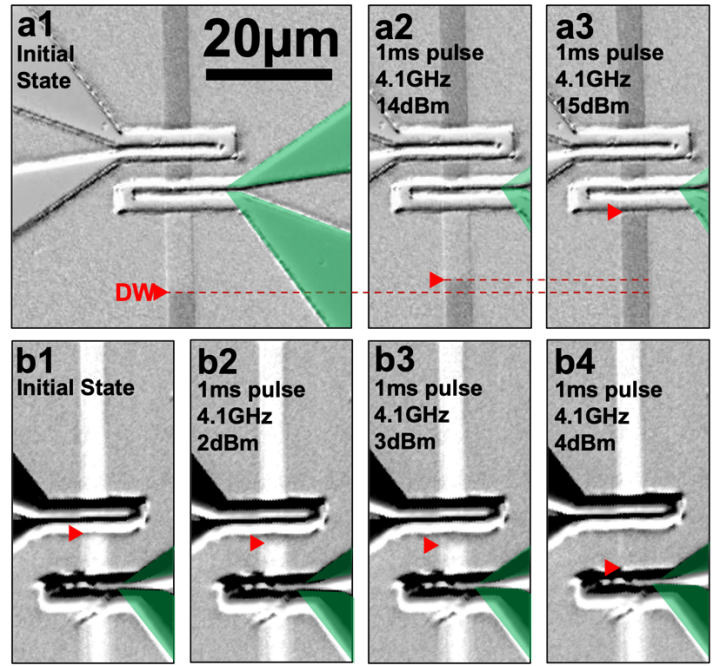

**Fig. 3 | Magnon-driven DW motion.** The injector antenna is highlighted in green and the DW is indicated with a red arrow. **a**, MOKE images of a DW in the initial state (a1), after the first pulse at 14 dBm (a2), and the second pulse at 15 dBm (a3) applied at the lower antenna, using a 1 ms pulse at 4.1 GHz. **b**, a DW formed near the upper antenna edge in the initial state (b1), after a 1 ms pulse at 2 dBm (b2), 3 dBm (b3), and 4 dBm (b4) applied at the lower antenna at 4.1 GHz.

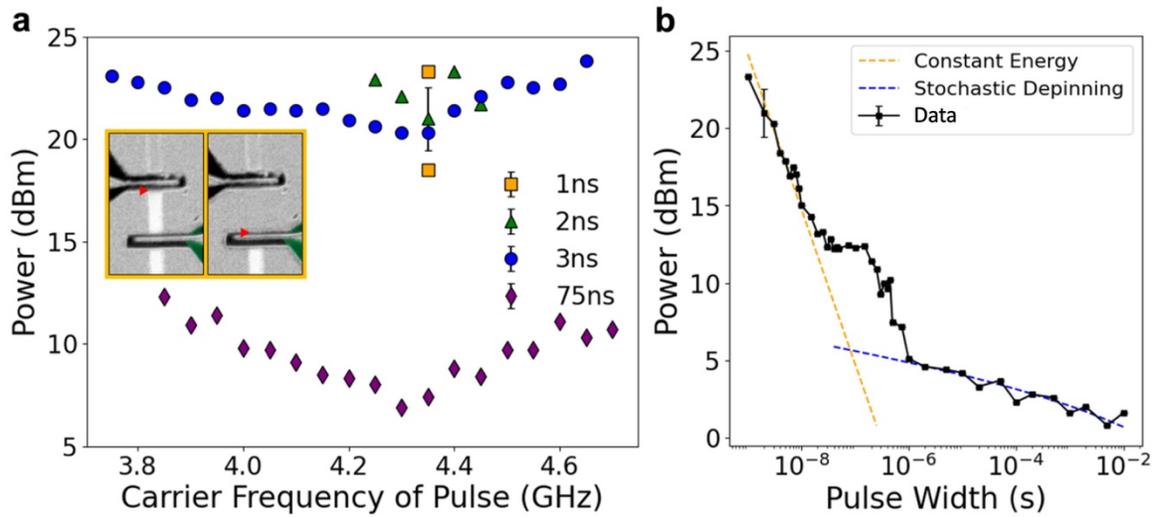

**Fig. 4 | Magnon-induced DW motion power threshold dependence on pulse width and carrier frequency of pulse. a**, Power threshold as a function of RF carrier frequency for pulse widths of 1, 2, 3, and 75 ns. Inset shows MOKE images of a 15 μm-gap device before and after the 1 ns 4.35 GHz pulse inducing DW motion at 19 dBm. **b,** DW motion power threshold as a function of pulse width with a carrier frequency of 4.35 GHz. The error bar is based on 10 repeated measurements. Orange dashed line shows the linear slope corresponding to a constant pulse energy. Blue dashed line corresponds to stochastic depinning fit for a pulse width ≥ 1 μs.